\documentclass{osa-article}
\usepackage{amsmath,amssymb,bm,color,graphicx}
\journal{osac}

\articletype{Research Article}

\begin{document}

\title{Observation of concentrating paraxial beams}

\author{Andrea~Aiello,\authormark{1} 
Martin~Pa\'{u}r,\authormark{2} Bohumil Stoklasa,\authormark{2}
Zden\v{ e}k~Hradil,\authormark{2} Jaroslav~\v{R}eh\'a\v{c}ek,\authormark{2}
and Luis~L.~S\'{a}nchez-Soto\authormark{1,3,*}}

\address{\authormark{1}Max-Planck-Institut f\"ur  die Physik des Lichts, 
Staudtstra\ss e 2, 91058 Erlangen, Germany \\
\authormark{2}Department of Optics, Palack\'{y}  University, 
17.~listopadu 12, 77146 Olomouc, Czech Republic\\  
\authormark{3}Departamento de \'{O}ptica, Facultad de F\'{\i}sica, 
Universidad Complutense, 28040~Madrid, Spain} 

\email{\authormark{*}lsanchez@fis.ucm.es} 



\begin{abstract}
We report,to the best of our knowledge, the first observation of concentrating paraxial beams of light in a linear nondispersive medium. We have generated this intriguing class of light beams, recently predicted by one of us, in both one- and two-dimensional configurations. As we demonstrate in our experiments, these concentrating beams display unconventional features, such as the ability to strongly focus in the focal spot of a thin lens like a plane wave, while keeping their total energy finite.
\end{abstract}

\section{Introduction}

Legend says that during the Siege of Syracuse the Greek mathematician Archimedes set the Roman fleet on fire using mirrors that concentrated sunlight upon the ships. Modern versions of Archimedes mirrors are solar furnaces that can achieve light intensities of about $10$~MW/m$^2$ at the focus.  

Focusing light is the province of mirrors and lenses.  A converging lens focuses a collimated beam onto a bright spot centered about the focal point, at a distance $f$ from the lens. Reversing the direction of light, the same lens can be used to convert a small spotlike source at the focus in a collimated beam. Therefore, if such a source covered an arbitrarily small area $d^2$, we would expect that in the reversed operation a converging lens could focus a collimated light beam on a spot of the same area~$d^2$.

However, things are not quite so simple. As a matter of fact, in 1873 Abbe found that a lens cannot focus light of wavelength $\lambda$ to a spot of area smaller then $\lambda^2$~\cite{Abbe:1873aa}. Since then, this result has been a tenet of wave optics and it is often presented as a fundamental constraint arising from the properties of the Fourier transform, which relates the transverse spatial spread of a wavepacket to its angular spread in the corresponding wavevector space~\cite{Wong:2017aa}.

Only recently various techniques aiming at beating this limit were proposed and eventually demonstrated~\cite{Huszka:2019aa}.  In particular, Pendry suggested that the physical origin of such $d^{2} > \lambda^{2}$ limitation resides in the impossibility of a lens to act upon those evanescent waves that would contribute to the formation of a pointlike spot in the focal plane~\cite{Pendry:2000aa}. To overcome this problem Pendry proposed the use of a suitably engineered \emph{superlens} that would amplify evanescent waves.

Here, we follow a completely different approach based on  a very  elementary observation: if evanescent waves constitute the origin of the problem, then it would be sufficient to consider only light waves that do not contain them. Paraxial optics deals precisely with such a kind of waves~\cite{Goodman:2005aa}. That a paraxial plane wave can be concentrated into a pointlike spot of infinite brightness by a thin lens, is a textbook result~\cite{Born:2003aa}. However, a \emph{bona fide} plane wave has an infinite transverse extension and convey infinite energy. Plane waves are not the only ones that require an infinite amount of energy to be generated:  Bessel~\cite{Durnin:1987aa} and Airy~\cite{Siviloglou:2007aa} beams are very famous examples thereof. Therefore, plane waves, Bessel beams and Airy beams, just to name a few popular nondiffracting beams, are abstractions which can be implemented only approximately in the real world.

In this work, we report the first observation of concentrating paraxial beams of light focused by a thin lens. As shown in Ref.~\cite{Aiello:2016aa}, these beams have an infinite lateral extent like, e.g., plane waves or Bessel beams, but they carry finite energy. Of course, in practice, these beams are typically truncated by an aperture. However, if the maximal transverse size of such aperture falls behind the focal length of the lens, paraxial regime of propagation is maintained and the truncated beams are still strongly concentrated in the focal point of the lens, thus preserving the major characteristics of ideal concentrating beams. In fact, our experiments demonstrate that even though the beams are truncated they still greatly outperform standard Gaussian beams in terms of focalized intensity.

\begin{figure}[t]
    \centering
    \includegraphics[width=.80\columnwidth]{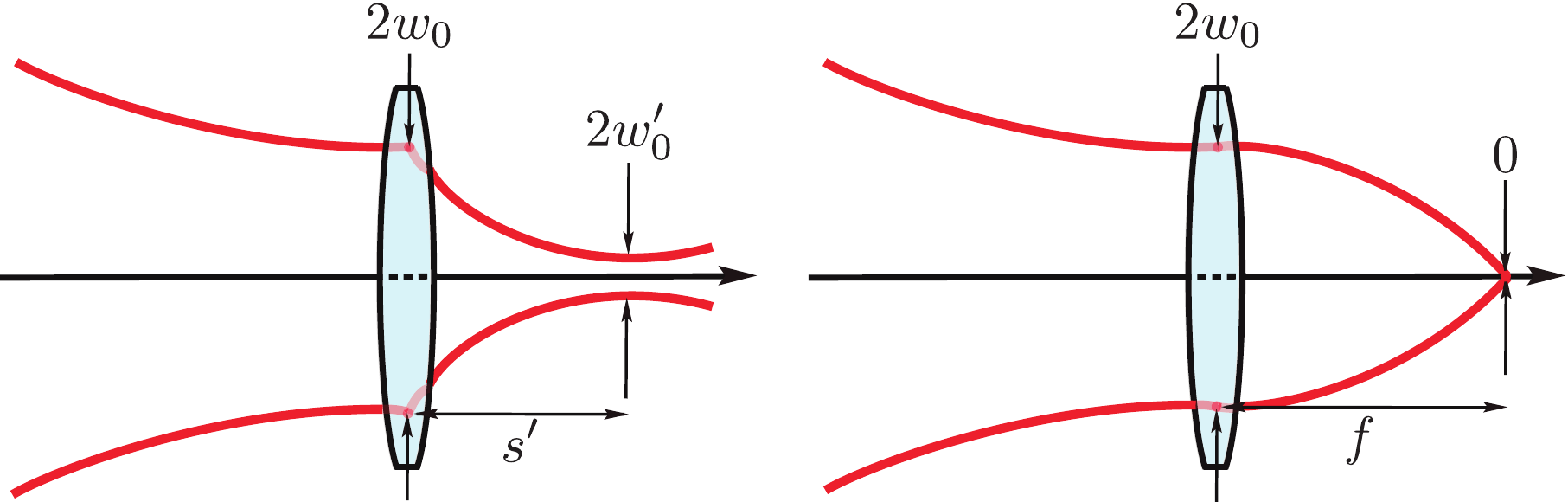}
    \caption{Illustration of the focusing behavior of a Gaussian beam
      (left) and a concentrating beam (right) after passing through a thin
      lenses.}
    \label{fig:foc}
\end{figure}

Paraxial optics is often erroneously considered synonymous of Gaussian optics. In this paper, we challenge this preconception with these concentrating beams, which do not follow the rules of Gaussian beam optics. This concept is exemplified in Fig.~\ref{fig:foc}, where the focusing properties of a Gaussian and a concentrating beam are compared. The minimum waist $w_0^{\prime}$ of a Gaussian beam of waist $w_0$ focused by a thin lens of focal length $f$ is
\begin{equation}
\frac{w_0^{\prime}}{w_0} = \frac{1}{\sqrt{1+z_R^2/f^2}} \, ,
\end{equation} 
and this occurs at a distance 
\begin{equation}
s^{\prime} = \frac{f}{1+f^{2}/z_{\mathrm{R}}^{2}}
\end{equation} 
from the lens, with $z_{\mathrm{R}}$ being the Rayleigh range~\cite{Almeida:1999aa}. This shows that $w_0^{\prime}/w_0 \to 0$ only when the focal length $f$ goes to zero and then $s^{\prime} \to 0$. This explains why extreme optical microscopy requires small focal length and samples close to the microscope, thus breaking the paraxial conditions. As we will see, for a concentrating beam the situation is the opposite: a zero-width waist is obtained at distance $f$ of the lens, where $f$ can be arbitrarily large.

Therefore, the new message we wish to convey is that, surprisingly, physically realizable paraxial waves that carry a \emph{finite} amount of energy and that, notwithstanding, can be focused onto a spot of (theoretical) infinite brightness, do exist.

Focusing optical beams has always been an issue of great practical importance. However, we stress that the significance of our results goes beyond the realm of optics. Actually, in quantum mechanics it is not difficult to construct states that are represented, at $t=0$, by a continuous wave function and then evolve (under the Schr\"{o}dinger equation) into a singular one~\cite{Peres:1994aa}. In view of the exact equivalence between the paraxial wave equation and the Schr\"{o}dinger equation~\cite{Yariv:1995aa}, our results can be seen as an experimental confirmation of such a singular blow-up, that has never been observed thus far. 

\section{Concentrating beams: Ideal case}

To begin with, let us set the stage for our simple theory.  We consider an electromagnetic wave of frequency $\omega$, which in the scalar approximation is described by its spatial amplitude ${U}(\mathbf{r})$.  We denote by $U(x, y)$ the value of this amplitude at the plane $z=0$, wherein the field has a Cartesian symmetry, so it factorizes as $U(x,y) = \mathfrak{a}(x) \mathfrak{a}(y)$. An ideal thin lens of focal length $f$ and transmission function  
\begin{equation} 
t_{\mathrm{lens}} (x, y) = \exp [ - {i kx^{2}}/{(2f)} ] 
\; \exp [ - {i ky^{2}}/{(2f)} ]
\end{equation} 
is placed at $z=0$. Using the angular spectrum representation~\cite{Mandel:1995aa}, the amplitude at an arbitrary plane $z>0$ is 
\begin{align}
\label{fieldz}
  U(x,y,z) =  \iint\limits_{-\infty}^{\quad + \infty} 
  u(p,q) \, e^{i k ( px + q y)} \, \, e^{i k \sqrt{1-p^2 -q^2} \, z }
  \, dp dq \, ,
\end{align} 
where
\begin{align}
\label{fieldFT}
  u (p,q) = & \; \frac{k^2}{(2 \pi)^2}
 \iint\limits_{-\infty}^{\quad + \infty} 
 t_{\mathrm{lens}} (x, y) \, U(x,y)
e^{-i k ( px + q y )} \, dx dy  
\end{align}
is the Fourier transform of the field at plane $z=0$ immediately after the lens. 

In the paraxial regime waves are characterized by small angular deviations with respect to the mean propagation axis, say $z$. In such a case, $p^2 + q^2 \ll 1$ and, to a good approximation, we can replace the unpleasant exact transfer function $\exp(i k z \sqrt{1-p^2 - q^2})$ with the friendly Fresnel propagator  $\exp \{  i k z  [ 1- (p^2+q^2)/2] \}$. This gives
\begin{align}
\label{fieldparz}
  U(x,y,z) =  \iint\limits_{-\infty}^{\quad + \infty} 
  u(p,q) \, e^{i k ( px + q y)} \,  
  e^{i k z \left[ 1 - (p^{2} + q^{2})/2 \right]}
  \, dp dq \, .
\end{align} 

To check this expression, let us first consider the somewhat simpler example of a plane wave of unit amplitude incident along the $z$ axis. The spectrum $u(p,q)$ can be now readily calculated; the final result reads $U_{\mathrm{pl}}(x,y, z) = 
\mathfrak{a}_{\mathrm{pl}} (x,z) \;  \mathfrak{a}_{\mathrm{pl}} (y,z)$,
with 
\begin{equation}
\label{eq:fieldzB}
 \mathfrak{a}_{\mathrm{pl}} (x,z) = \frac{ e^{i k z}}
 {\sqrt{2 \pi i kf}} \int\limits_{-\infty}^{\quad + \infty}
  \exp\left[i \left(  p^2\frac{1- z/f}{2 kf} \,  + 
  p \frac{x}{f} \right)\right] \, d p \, ,
\end{equation}
and an analogous expression for $\mathfrak{a}_{\mathrm{pl}} (y,z)$. When the lens is absent,  $u(p,q) = \delta(p) \delta(q) $ and this yields $U(x, y, z) = \exp(i k z)$, as it should be. In the focal plane of the lens, ${z} = f$, and Eq.~\eqref{eq:fieldzB} gives at once
\begin{align}
  \label{fieldzC}
 U(x, y, z=f) = \frac{2 \pi e^{2 i k f}}{i kf} \,   
 \delta (x/f) \, \delta (y/f) \, ,
\end{align}
which  means that, as one might anticipated, a plane wave is focused by a thin lens into a pointlike spot of infinite brightness.

\begin{figure}[t]
  \centering 
  \includegraphics[width=0.75\columnwidth]{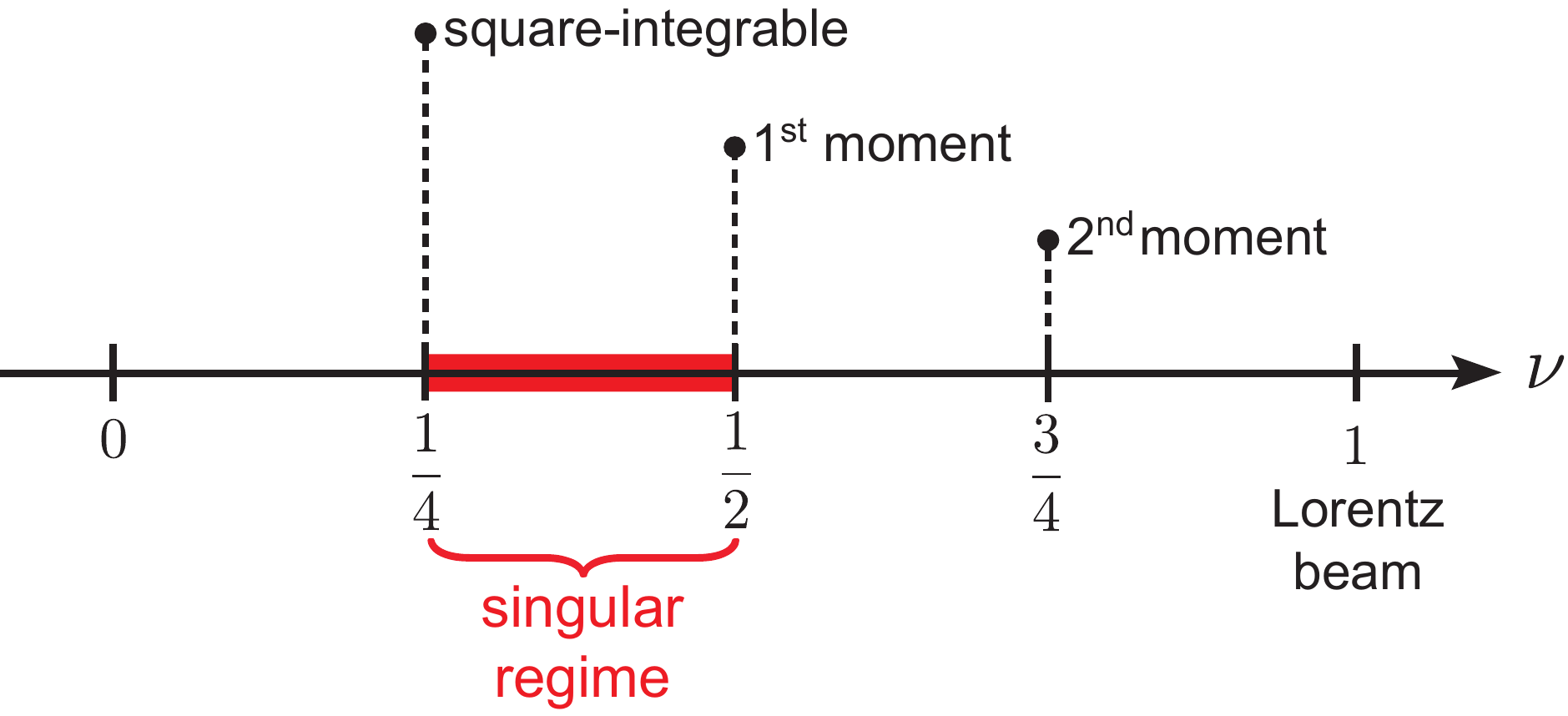}
  \caption{For $1/4 < \nu$ the function $\mathfrak{a}_\nu (x,0)$ is square integrable. For $1/4 < \nu < 1/2$ (represented as a red band) $\mathfrak{a}_\nu (x,z)$ becomes infinite at the focal point $z=f$. For $1/2 < \nu$, $\mathfrak{a}_\nu (x,z)$ is finite everywhere and the first moment $\langle x \rangle$ of the intensity distribution $|\mathfrak{a}_\nu (x,z) |^2$ exists and it is equal to $0$. Finally, the second moment $\langle x^2 \rangle$ is finite for $3/4 < \nu$ and gives a faithful measure of the beam waist. For $\nu = 1$ the concentrating beam becomes a so-called Lorentz beam~\cite{Gawhary06}.}
   \label{fig:Range_nu}
\end{figure}

The ideal focusing of a plane wave is well understood, but a plane wave is not a normalizable field and thus is not physically feasible. Our goal is to study a family of paraxial fields, square integrable and almost everywhere smooth, that spontaneously generate a singularity along the propagation axis in the focal point of a thin lens. We will refer to them as concentrating beams. According to the ideas developed in Ref.~\cite{Aiello:2016aa} we consider the family (up to a global constant)
$U_{\nu} (x,y) = \mathfrak{a}_{\nu} (x) \, \mathfrak{a}_{\nu} (y)$, with 
\begin{equation}
\label{eq:Using}
\mathfrak{a}_{\nu} (x) = \frac{1}{(1 + x^{2}/\sigma_{\nu}^{2})^{\nu}} \, ,
\end{equation}
where  $\sigma_{\nu}  > 0$ determines the spot size of the beam and $\nu > 0$ is a positive constant that can be varied to obtain different kind of singularities. For $\nu = 1$ they reduce to the so-called Lorentz beam~\cite{Gawhary06}. A similar expression holds for $\mathfrak{a}_{\nu} (y)$.  Using again the paraxial approximation \eqref{fieldparz}, we obtain $U_{\nu} (x, y, z) = {k}/{(2 \pi i z)} \;  \mathfrak{a}_{\nu} (x,z ) \, \mathfrak{a}_{\nu} (y, z)$, where now
\begin{equation}
\mathfrak{a}_{\nu} (x,z ) = \sqrt{\frac{k}{2 \pi i z}}
e^{i \frac{k x^{2}}{2z}} 
\int_{-\infty}^{+\infty} \frac{\exp \left \{ i \frac{k}{2z} 
\left [ - 2 x  s + (1 - z/f) s^2 \right ]\right \}}
{(1+ s^{2}/\sigma_{\nu}^{2})^{\nu}} \, ds \, .
\end{equation}
One can check that for $1/4 < \nu < 1/2$ the beams (\ref{eq:Using}) are square integrable
\begin{align}
  \label{Lorentz2}
\iint\limits_{-\infty}^{\quad + \infty}   
| U_{\nu} (x, y) |^2 \, dxdy =  \pi  \sigma_{\nu} 
\left [ \frac{\Gamma\left(2 \nu - 1/2 \right)}
  {\Gamma(2 \nu)} \right ]^{2} \, ,
\end{align}
but the field $U_{\nu}(x,y,z)$ diverges at the focal point of the lens $z=f$. 

\section{Concentrating beams: realistic case}
The theory thus far truly confirms that, in principle, it is possible to achieve an infinitely bright spot by focusing a nonplane wavefront with an ideal thin lens. However, this is a speculative situation, for any lens always has a finite aperture, which causes that the amplitude of our beam remains finite at $z=f$. In addition, we have ignored any polarization effect. 

To cope with these problems, we take the monomochromatic field at $z=0$ as the vector field $\mathbf{E} ( x, y) = U(x, y) \, \hat{\mathbf{e}}_{p} (\alpha, \beta)$, where $U(x, y)$ is the spatial amplitude and $\hat{\mathbf{e}}_{p} (\alpha, \beta) = \hat{\mathbf{e}}_{x} \cos \alpha + e^{i \beta} \hat{\mathbf{e}}_{y} \sin \alpha$ is the unit polarization vector, with direction angles $(\alpha, \beta)$.  We take the lens as having an aperture $2L$ in each direction. The field at a plane $z$ can be exactly computed according to
\begin{equation}
\label{eq:vecex}
\mathbf{E} (x,y,z) = \int_{-L}^{L} \int_{-L}^{L} 
U (\xi, \eta) \, t_{\mathrm{lens}} (\xi, \eta) \;
\mathbf{K} (x-\xi, y- \eta, z) \, d\xi d\eta \, ,
\end{equation}
where 
\begin{eqnarray}
\mathbf{K} ( x-\xi, y - \eta, z )  = \frac{1}{2\pi}
\left \{  - \hat{\mathbf{e}}_{p} 
\frac{\partial G (\mathbf{r} - \mathbf{r}_{0})}{\partial z} 
+ \hat{\mathbf{e}}_{z} 
\left[ \hat{\mathbf{e}}_{p} \cdot \hat{\mathbf{e}}_{x} 
\frac{\partial G (\mathbf{r} - \mathbf{r}_{0})}{\partial x} 
+ \hat{\mathbf{e}}_{p} \cdot \hat{\mathbf{e}}_{y} 
\frac{\partial G(\mathbf{r} - \mathbf{r}_{0})}{\partial y} \right ]  \right \} \,. &
\end{eqnarray}
Here $(\xi, \eta)$ are the coordinates of a point in the aperture of the lens, $G (\mathbf{r} - \mathbf{r}_{0} ) = {\exp (i k |\mathbf{r} - \mathbf{r}_{0}|)}$$/{|\mathbf{r} - \mathbf{r}_{0}|}$ is the Green function of the problem and $\mathbf{r} - \mathbf{r}_{0} =   \hat{\mathbf{e}}_{x} (x-\xi) + \hat{\mathbf{e}}_{y} (y-\eta) + \hat{\mathbf{e}}_{z} z$.  In this way, $\mathbf{E} (x,y, z)$ verifies the Helmoltz equation, and not only the paraxial approximation, in every point of the space, as one can easily check.

\begin{figure*}[t]
    \centering{\includegraphics[width=\columnwidth]{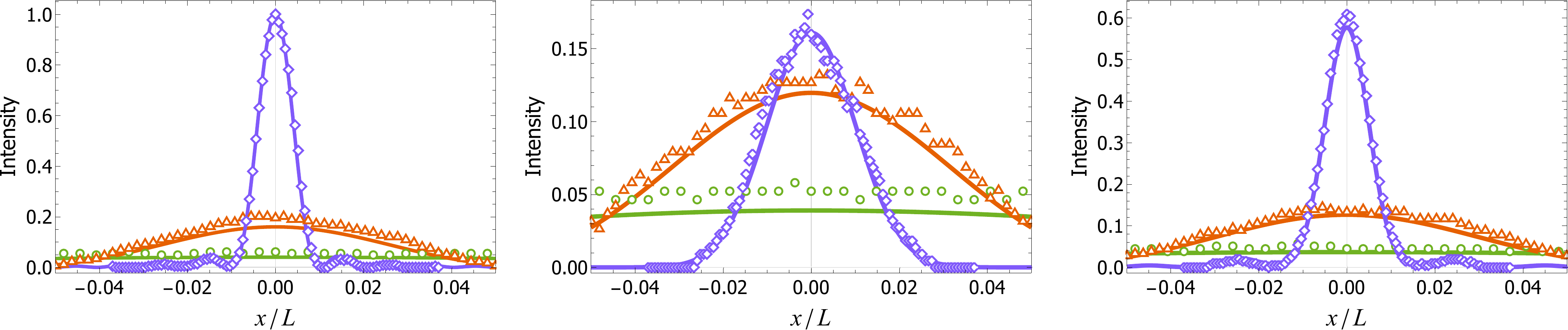}}
    \caption{Comparison of the intensity profiles $|\mathbf{E}(x,0,f)|^{2}$ at the focal point of the lens $z=f$. The vertical scale is fixed by the normalization condition~(\ref{fields}) for $L=2.5$~mm.  Left panel corresponds to plane waves, central panel to Gaussian beams, and right panel to concentrating beams with $\nu = 1/3$. In each plot, we represent the results for three different apertures of $L=0.5$~mm (green), 1~mm (orange) and 2.5~mm (blue). The continuous lines correspond to the theoretical values, whereas the points indicate experimental values. The input polarization is always linear and parallel to the $y$ axis.}
    \label{fig:comp}
\end{figure*}
\begin{figure}[b]
    \centering
    \includegraphics[width=0.55\columnwidth]{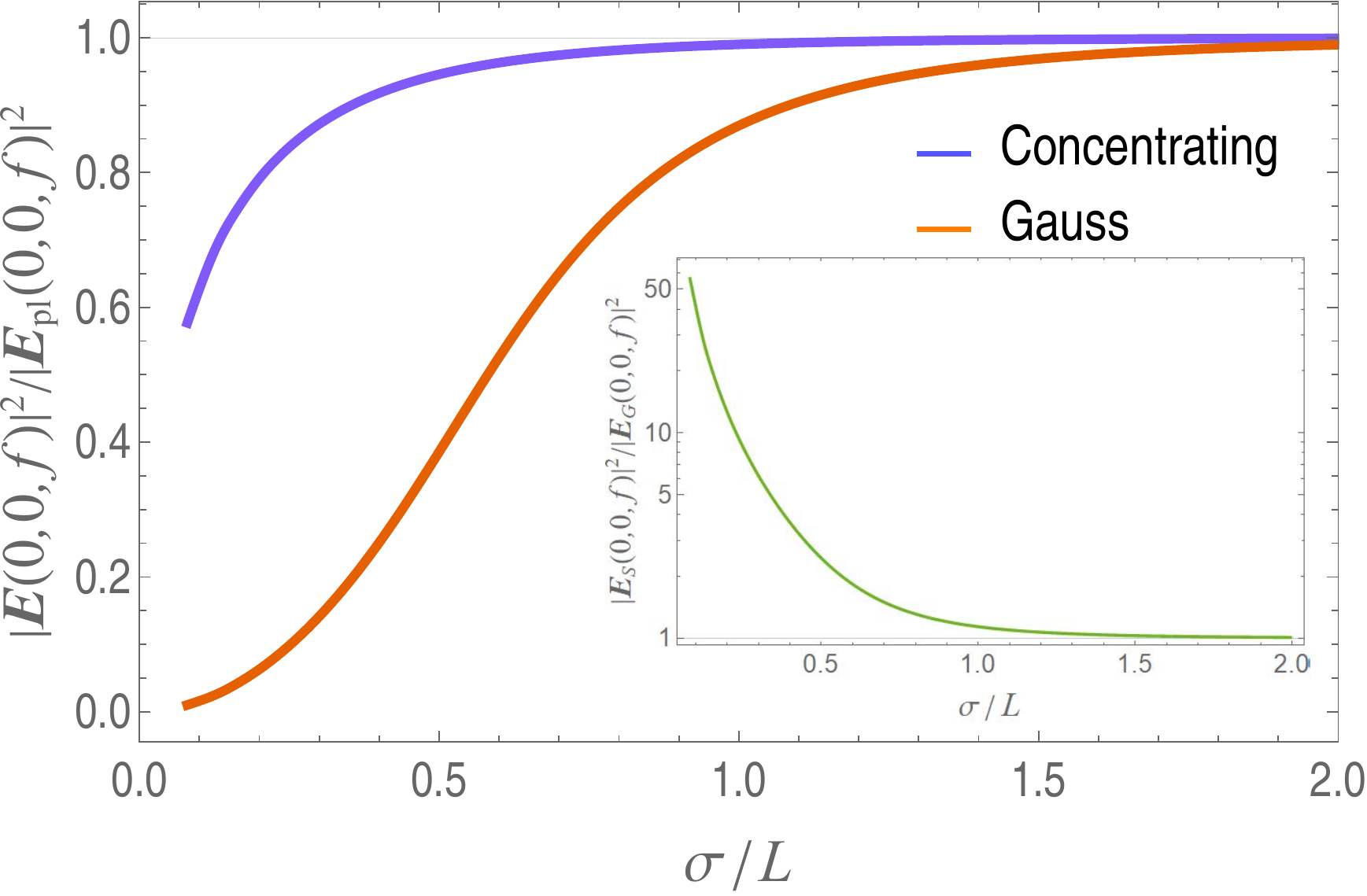}
    \caption{On-axis intensities at the focal plane $z=f$ for concentrating and Gaussian beams (normalized to the corresponding value for a plane wave) as a function of their dimensionless widths (in units of $L$). In the inset, we show the ratio of the on-axis intensities of the concentrating to the Gaussian beam in terms of the same parameter.}
    \label{fig:asym}
\end{figure}
To gain insight into the problem, we will compare the focusing performance of a plane wave and a singular Lorentz beam, which henceforth we will take with $\nu=1/3$. For completeness, we also include a Gaussian beam, since this is perhaps the most prevalent of all beams. To make a fair comparison we normalize these fields so as they have unit intensity across the aperture; that is, 
\begin{equation}
\label{eq:norm}
\int_{-L}^{L} \int_{-L}^{L} \mathbf{E}^{\ast} (x, y, 0) 
\cdot \mathbf{E} (x, y, 0) \, dx dy = \int_{-L}^{L} \int_{-L}^{L} \ 
| U (x, y) |^{2} \, dx dy = 1 \, ,
\end{equation} 
which fixes the field amplitudes to the following forms
\begin{align}
   \label{fields}
   U_{\mathrm{pl}} (x, y) & = \frac{1}{2L} \, ,\nonumber \\
U_{\nu}(x,y) & = \frac{1}{2L} 
\frac{(1 + x^{2}/\sigma_{\nu}^{2})^{- \nu} (1 + y^{2}/\sigma_{\nu}^{2})^{- \nu}
}{\/_2F_1 ( 1/2, \nu, 3/2; -L^2/\sigma_{\nu}^{2} ) } \,, \\
U_{\mathrm{G}} (x,y) & = \frac{\sqrt{2}}{\sqrt{\pi} \, \sigma_{\mathrm{G}} \, \mathrm{erf}(\sqrt{2} L/\sigma_{\mathrm{G}})}  \exp [ - (x^{2} + y^{2})/\sigma_{\mathrm{G}}^{2}] \, , \nonumber
\end{align}
$\/_2F_1 (a, b, c;z)$ and $\mathrm{erf} (z)$ being the hypergeometric and error functions, respectively. 

We have numerically integrated the exact result in Eq.~\eqref{eq:vecex} for these fields and three different values of the aperture: $L=0.5, 1$ and 2.5~mm. The resulting intensity profiles at the focal point of the lens are shown in Fig.~\ref{fig:comp}. In all the cases we have chosen linear polarization parallel to the $y$ axis. At $z=f$, the polarization is still linear without any appreciable $z$ component, which confirms that polarization plays no significant role in the behavior of the singularity. We can appreciate that the size of the Gaussian spot remains approximately constant with the aperture, whereas the concentrating field reveals a clear focusing when the aperture increases, and the same happens for the plane wave.

When the width of both concentrating and Gaussian beams exceed $L$, the corresponding intensity profiles approach the plane wave (uniform) intensity distribution. The maximum value corresponding to the plane wave is obtained asymptotically for both beams, but the concentrating beam approaches it much faster. This is confirmed in Fig.~\ref{fig:asym}, where we plot the maximum intensity at the focus $z=f$ for both beams (normalized to the value of the plane wave).

\begin{figure}[b]
    \centering
    \includegraphics[width=0.75\columnwidth]{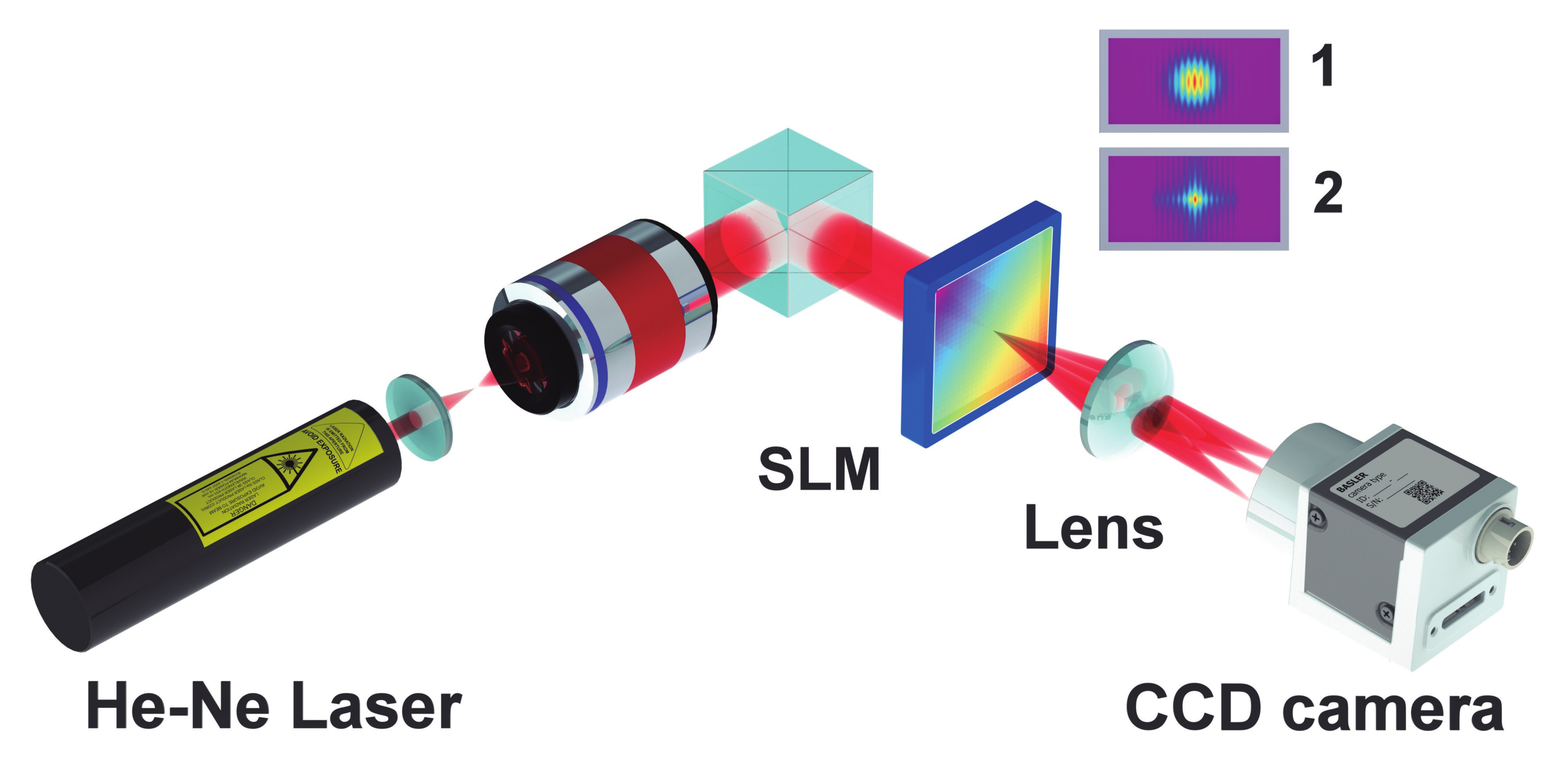}
    \caption{Schematic of the experimental setup for the generation of
      concentrating beams. See the description in the text for more details.}
    \label{fig:setup_L}
\end{figure}

\section{Experimental results}

We have checked these predictions in the laboratory (Fig.~\ref{fig:setup_L}.  To build up the desired beam profile we used an amplitude spatial light modulator (SLM) (CRL OPTO) illuminated by a collimated beam from linearly polarized He-Ne laser (633~nm Thorlabs). The modulator was calibrated to s transmitted wavefront with the final Strehl ratio better than 0.98. Next, calibration of the amplitude modulation linearity was performed, from where we  estimate our systematic errors.  On the SLM, several holograms with different aperture sizes and shapes were realized. These holograms were generated as an interference between a tilted reference plane wave and the function of desired beam profile. The beam complex amplitude is carried by the first diffraction order. This diffraction order was then Fourier transformed by a spherical lens of focal length $f=200$~mm and the signal was detected by CMOS camera (Basler) with $1.85 \times 1.85~\mu$m$^{2}$-pixel size.   

\begin{figure}[t]
    \centering
    \includegraphics[width=0.55\columnwidth]{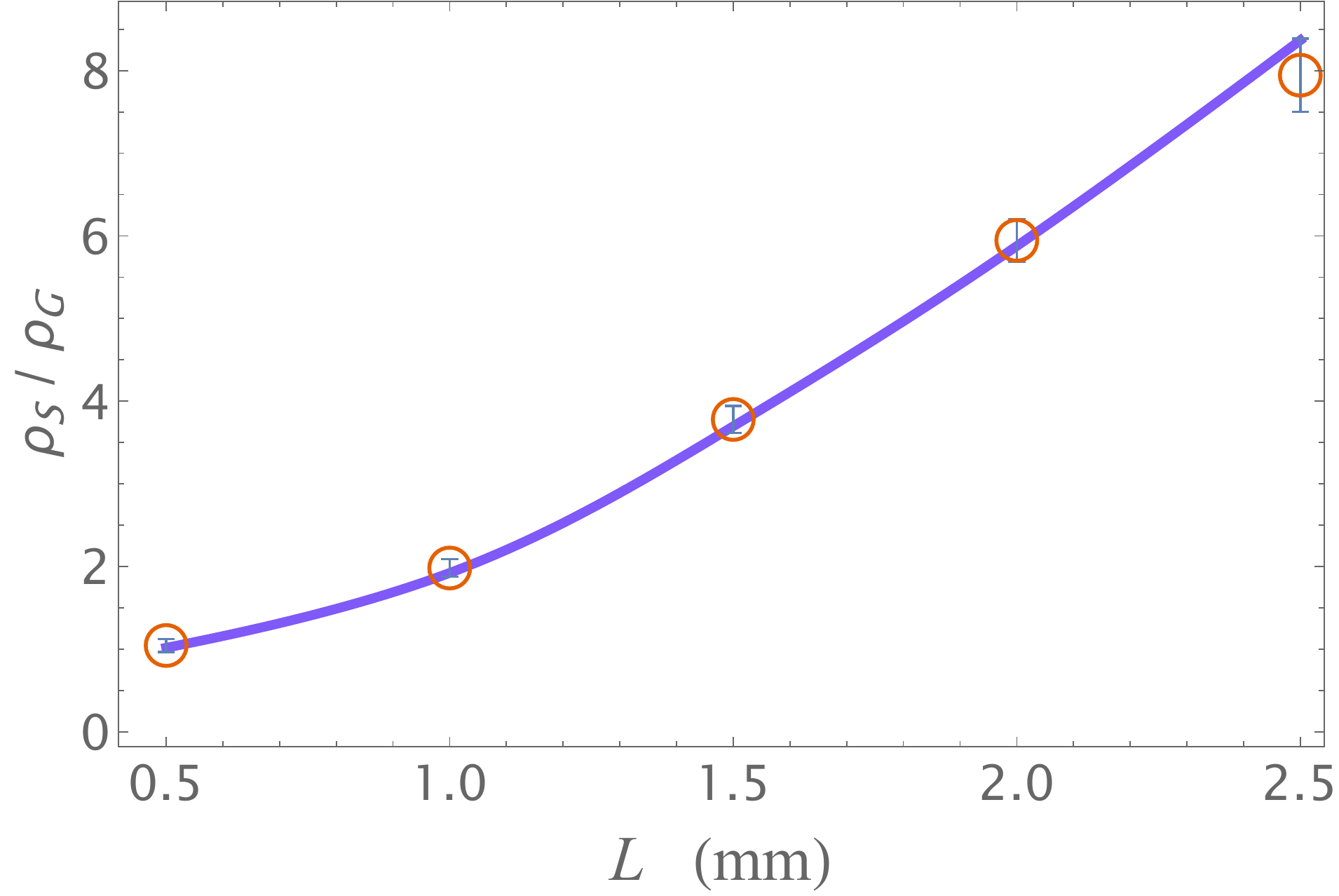}
    \caption{Ratio of the on-axis intensities $\varrho_{S}/\varrho_{G}$ of the concentrating to Gaussian beams, as the aperture increases. The experimental points are marked by red error bars (the error being estimated from the calibration), corresponding to $L = 0.5$, 1, 1.5, 2, and 2.5~mm. The red circles surrounding the error bars serve to help visualization.}
    \label{fig:ratio}
\end{figure}

In this way, we reproduced the theoretical predictions of Fig.~\ref{fig:comp}. This required to select the widths $\sigma_{\nu}$ and $\sigma_{G}$ for each aperture individually. The experimental results are in excellent agreement with the theory.

\begin{figure}[b]
  \centering 
  \includegraphics[width=.95\columnwidth]{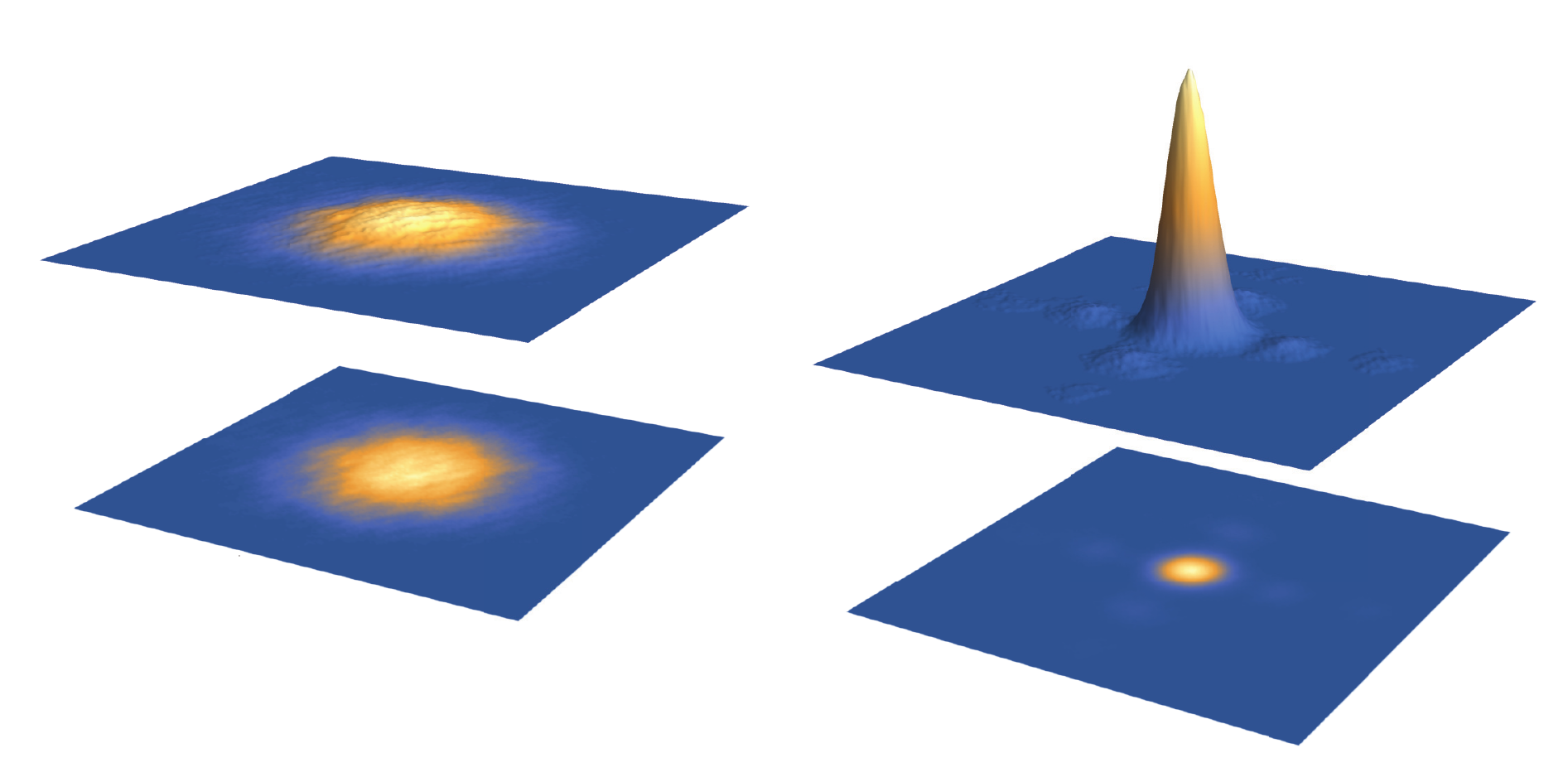}
  \caption{Raw intensity data for the concentrating beam (right) and the Gaussian beam (left) for the aperture of 2.5~mm. At the bottom, we plot density plots of the intensities as seen in the CMOS camera at the focal spot of the lens. The (arbitrary) unit of the vertical scale is the same for both plots.}
  \label{fig:data2D}
\end{figure}

Experimental data shows an important increase of on-axis intensity of the concentrating beam when the energy of the beam is slightly increased. For this reason, we measured the ratio $\varrho_{S}/\varrho_{G}$ of the on-axis intensities of the concentrating to the Gaussian beam, as the aperture increases, but keeping the corresponding beam widths:  $\sigma_{\nu} = 0.16$  and $\sigma_{\mathrm{G}} = 0.5$. The results are presented in Fig.~\ref{fig:ratio}, which brings out the advantages of the concentrating beam in comparison with its Gaussian counterpart. 

Finally, in Fig.~\ref{fig:data2D} we include raw intensity data at the focal point from this last experiment for $L= 2.5$~mm. The real camera pictures provide a dramatic evidence of the focusing capabilities of the concentrating beams.

To sum up, we have presented and experimentally realized a family of square-integrable paraxial beams that spontaneously develop a singularity when they propagate in free space.  We stress that this is a linear property of the beams and not the outcome of any self-focusing or other nonlinear effects~\cite{Efremidis:2010aa}.  While in the standard picture of extreme light focusing the minimum focal spot is achieved with large numerical apertures, in our new paradigm the opposite is true: strong focusing is achieved moving towards paraxial optics, as in the case of a flat (plane wave) illumination of an aperture. However, for a given fixed average input intensity across the aperture, the energy carried by a plane wave increases quadratically with the linear size of the aperture, whereas for our concentrating beams the energy remains constant. This promising field enhancement mechanism may foster further interesting research in classical and quantum optics. 

\noindent \textbf{Funding.} European Union Horizon 2020 (732894), 
Grant Agency of the Czech Republic (18-04291S), 
Palack{\'y} University (IGA\_PrF\_2020\_004), 
Ministerio de Ciencia e Innovaci\'on (PGC2018-099183-B-I00)

\noindent\textbf{Disclosures.} 
The authors declare no conflicts of interest.


\begin{thebibliography}{10}
\newcommand{\enquote}[1]{``#1''}

\bibitem{Abbe:1873aa}
E.~Abbe, \enquote{Beitr\"{a}ge zur theorie des mikroskops und der
  mikroskopischen wahrnehmung,} {\protect\JournalTitle{Arch. Mikrosk. Anat.}}
  \textbf{9}, 413--468 (1873).

\bibitem{Wong:2017aa}
L.~J. Wong and I.~Kaminer, \enquote{Abruptly focusing and defocusing needles of
  light and closed-form electromagnetic wavepackets,}
  {\protect\JournalTitle{ACS Photonics}} \textbf{4}, 1131--1137 (2017).

\bibitem{Huszka:2019aa}
G.~Huszka and M.~A. Gijs, \enquote{Super-resolution optical imaging: A
  comparison,} {\protect\JournalTitle{Micro Nano Engineering}} \textbf{2},
  7--28 (2019).

\bibitem{Pendry:2000aa}
J.~B. Pendry, \enquote{Negative refraction makes a perfect lens,}
  {\protect\JournalTitle{Phys. Rev. Lett.}} \textbf{85}, 3966--3969 (2000).

\bibitem{Goodman:2005aa}
J.~W. Goodman, \emph{Introduction to Fourier Optics} (Roberts \& Company,
  Englewood, CO, 2005), 3rd ed.

\bibitem{Born:2003aa}
M.~Born and E.~Wolf, \emph{Principles of Optics} (Cambridge University Press,
  Cambridge, 2003), 7th ed.

\bibitem{Durnin:1987aa}
J.~Durnin, J.~J. Miceli, and J.~H. Eberly, \enquote{Diffraction-free beams,}
  {\protect\JournalTitle{Phys. Rev. Lett.}} \textbf{58}, 1499--1501 (1987).

\bibitem{Siviloglou:2007aa}
G.~A. Siviloglou, J.~Broky, A.~Dogariu, and D.~N. Christodoulides,
  \enquote{Observation of accelerating {A}iry beams,}
  {\protect\JournalTitle{Phys. Rev. Lett.}} \textbf{99}, 213901 (2007).

\bibitem{Aiello:2016aa}
A.~Aiello, \enquote{Spontaneous generation of singularities in paraxial optical
  fields,} {\protect\JournalTitle{Opt. Lett.}} \textbf{41}, 1668--1671 (2016).

\bibitem{Almeida:1999aa}
A.~M. Almeida, E.~Nogueira, and M.~Belsley, \enquote{Paraxial imaging: Gaussian
  beams versus paraxial-spherical waves,} {\protect\JournalTitle{Am. J. Phys.}}
  \textbf{67}, 428--433 (1999).

\bibitem{Peres:1994aa}
A.~Peres, \emph{Quantum theory: Concepts and methods} (Kluver, San Diego, CA,
  1994), 5th ed.

\bibitem{Yariv:1995aa}
A.~Yariv, \emph{Optical Electronics} (Cambridge University Press, New York,
  1995), 4th ed.

\bibitem{Mandel:1995aa}
L.~Mandel and E.~Wolf, \emph{Optical {C}oherence and {Q}uantum {O}ptics}
  (Cambridge University Press, New York, 1995).

\bibitem{Gawhary06}
O.~E. Gawhary and S.~Severini, \enquote{{Lorentz beams and symmetry properties
  in paraxial optics},} {\protect\JournalTitle{J. Opt. A: Pure Appl. Opt.}}
  \textbf{8}, 409--414 (2006).

\bibitem{Efremidis:2010aa}
N.~K. Efremidis and D.~N. Christodoulides, \enquote{Abruptly autofocusing
  waves,} {\protect\JournalTitle{Opt. Lett.}} \textbf{35}, 4045--4047 (2010).

\end{thebibliography}

\end{document}